\documentclass[english]{article}
\usepackage[T1]{fontenc}
\usepackage[latin9]{inputenc}
\usepackage{amsmath}
\usepackage{amssymb}
\usepackage{graphicx}
\usepackage{esint}
\usepackage{babel}
\begin{document}

\title{Electric and Magnetic fields due to Dirac particles in FRW spacetime}

\author{S. K. Sharma, P.R. Dhungel and U. Khanal \\
 Central Department of Physics, Tribhuvan University, Kirtipur, \textbf{Nepal}}
\maketitle
\begin{abstract}
Some solutions of the Maxwell equations with Dirac particles for the
source in FRW spacetime are discussed. The Green's function of the
equation for the radial component of the Maxwell fields, $F_{r\eta}$
and $F_{\theta\phi}$is solved. Green's function is found to reduce
to that of Minkowskian spacetime in the appropriate limit. Also, the
Lienard-Wiechert type solution is derived. Also, the solutions with
the Dirac particle current is also presented. It is found that the
$F_{r\eta}$is composed of even angular momentum states while the
odd states constitue $F_{\theta\phi}$.
\end{abstract}


\section{Introduction}

In Newmann-Penrose formalism \cite{Newman-Penrose-1962}, the six
components of the Maxwell field tensors are represented by three complex
scalars $\phi_{0}$, $\phi_{1}$and $\phi_{2}$. In the tetrad frame
we have chosen, \cite{Chandra-BH-1983,UK1} they are

\begin{equation}
\phi_{0}=F_{\mu\nu}l^{\mu}m^{\nu}=\frac{1}{\sqrt{2}a^{3}\sin r}\left[F_{r\theta}+F_{\eta\theta}+\frac{i}{\sin\theta}\left(F_{\eta\theta}+F_{r\phi}\right)\right]
\end{equation}
,

\begin{equation}
\phi_{1}=\frac{1}{2}F_{\mu\nu}\left(l^{\mu}n^{\nu}+\bar{m}^{\mu}m\nu\right)=\frac{1}{2a^{2}\sin^{2}r}\left[\sin^{2}rF_{r\eta}+\frac{i}{\sin\theta}F_{\theta\phi}\right]
\end{equation}

and 
\begin{equation}
\phi_{2}=F_{\mu\nu}\bar{m}^{\mu}n^{\nu}=\frac{1}{2\sqrt{2}a\sin r}\left[F_{r\theta}-F_{\eta\theta}+\frac{i}{\sin\theta}\left(F_{\eta\theta}-F_{r\phi}\right)\right]
\end{equation}
.

In projections of quantities in the tetrad frame, the number of $m$
minus the number of $\bar{m}$ is called the spin weight, while the
number of $\mathrm{\mathcal{\mathit{l}}}$ minus the number of $\mathrm{\mathcal{\mathit{n}}}$
is the boost weight. As a consequence, we see that $\phi_{1}$ has
spin and boost weight equal to zero and thus behaves as a spin zero
(scalar) field. Similarly, $\phi_{0}$ and $\phi_{2}$ are spin +1
and -1 fields.

Next, the Maxwell equations, with source by the Dirac current (\cite{UK1})
take on the form 

\begin{eqnarray}
\sin r\mathfrak{\mathtt{\mathcal{D_{\textrm{0}}^{\textrm{-}}\mathsf{\Phi_{0}}\textrm{\,\,-\,\,}L_{\textrm{1}}^{\textrm{+}}\mathsf{\Phi_{1}}}}} & = & \frac{\sin r}{\sqrt{2}}\left|\Phi_{\textrm{1/2}}\right|^{2}\left(\left|Y_{\textrm{1/2}}\right|^{2}+\left|Y_{-\textrm{1/2}}\right|^{2}\right)=J_{0}^{+}
\end{eqnarray}

\begin{eqnarray}
\sin r\mathfrak{\mathtt{\mathcal{D_{\textrm{-1}}^{\textrm{+}}\mathsf{\Phi_{1}}\textrm{+}L_{\textrm{0}}^{\textrm{-}}\mathsf{\Phi_{0}}}}} & = & \frac{\sin r}{\sqrt{2}}\left(\Phi_{\textrm{1/2}}^{*}\Phi_{\textrm{-1/2}}-\Phi_{\textrm{1/2}}\Phi_{-\textrm{1/2}}^{*}\right)Y_{\textrm{1/2}}Y_{-\textrm{1/2}}^{*}=J_{1}
\end{eqnarray}

\begin{eqnarray}
\sin r\mathfrak{\mathtt{\mathcal{D_{\textrm{-1}}^{\textrm{-}}\mathsf{\Phi_{-1}}\textrm{+}L_{\textrm{0}}^{\textrm{+}}\mathsf{\Phi_{0}}}}} & = & J_{1}^{*}=-J_{-1}
\end{eqnarray}

and
\begin{eqnarray}
\sin r\mathfrak{\mathtt{\mathcal{D_{\textrm{0}}^{\textrm{+}}\mathsf{\Phi_{0}}\textrm{\,\,\,-\,\,\,}L_{\textrm{1}}^{\textrm{-}}\mathsf{\Phi_{-1}}}}} & = & \frac{\sin r}{\sqrt{2}}\left|\Phi_{\textrm{-1/2}}\right|^{2}\left(\left|Y_{\textrm{1/2}}\right|^{2}+\left|Y_{-\textrm{1/2}}\right|^{2}\right)=J_{0}^{-}
\end{eqnarray}

where we have substituted $\Phi_{0}=\sqrt{2}a^{2}\sin^{2}r\Phi_{1}$,
$\Phi_{+1}=a^{3}\sin^{2}r\Phi_{0}$ and $\Phi_{-1}=2a\sin^{2}r\Phi_{2}$.
The radial-temporal operator $\mathfrak{\mathcal{D}}_{\textrm{s}}^{\pm}=\left(\frac{\partial}{\partial\theta}\mp\frac{i}{\sin\theta}\frac{\partial}{\partial\phi}+s\cot r\right)$and
the angular operator $\mathfrak{\mathcal{L}}_{\textrm{s}}^{\pm}=\left(\frac{\partial}{\partial\theta}\mp\frac{i}{\sin\theta}\frac{\partial}{\partial\phi}+s\cot\theta\right)$,
and $\eta$ is the conformal time $dt=ad\eta$ with $a$ equal to
the scale factor.

We can easily decouple the Maxwell equations to read \cite{PRD-UK} 

\begin{eqnarray}
\left[\sin r\mathsf{\mathsf{\mathcal{D}}}_{-1}^{\pm}\sin rD_{\textrm{0}}^{\mp}+\mathcal{L}_{\textrm{1}}^{\pm}\mathcal{L}_{\textrm{0}}^{\mp}\right]\Phi_{0} & = & \sin r\mathsf{\mathsf{\mathcal{D}}}_{-1}^{\pm}J_{0}^{\pm}+\mathcal{L}_{\textrm{1}}^{\pm}J_{\pm1}=S_{0}\label{eq:Maxwell Eq1}
\end{eqnarray}

and 
\begin{eqnarray}
\left[\sin r\mathsf{\mathsf{\mathcal{D}}}_{0}^{\mp}\sin rD_{\textrm{-1}}^{\pm}+\mathcal{L}_{\textrm{0}}^{\mp}\mathcal{L}_{\textrm{1}}^{\pm}\right]\Phi_{1} & = & \pm\sin r\mathsf{\mathsf{\mathcal{D}}}_{0}^{\mp}J_{\pm1}\mp\mathcal{L}_{\textrm{0}}^{\mp}J_{0}^{\pm}=S_{\pm1}\label{eq:Maxwell Eq2}
\end{eqnarray}

In Eq. (\ref{eq:Maxwell Eq2}), it is seen that $J_{0}$ is the source
for the radial electric field and $J_{1}$ for the radial magnetic
field. The eigenfunctions of the angular operator are the well known
spin-weighted spherical harmonics satisfying $\mathcal{L}_{\textrm{-\ensuremath{\left(s-1\right)}}}^{\mp}\mathcal{L}_{\textrm{s}}^{\pm}Y_{\pm}=-\left(l+s\right)\left(l-s+1\right)Y_{\pm}$
; we have identified these functions as the spherical harmonics formed
with the Jacobi polynomials \textit{viz.} 

\begin{equation}
Y_{\theta s}=_{s}Y_{l}^{m}=\frac{e^{im\phi}}{\sqrt{2\pi}}N\left(1-\cos\theta\right)^{\frac{m+s}{2}}\left(1+\cos\theta\right)^{\frac{m-s}{2}}P_{l-m}^{\left(m+s,m-s\right)}\left(\cos\theta\right)
\end{equation}

while $Y_{0}^{s}=Y_{l}^{m}$ are usual spherical harmonics. These
are normalized. So, $\int_{s}Y_{l_{1}\,\,\,\, s}^{m_{1}^{*}}Y_{l_{2}}^{m_{2}}d\Omega=\delta_{l_{1},l_{2}}\delta_{m_{1},m_{2}}$,
are complete in $\underset{l,m\,\,\,}{\sum_{\,\, s}}Y_{l}^{m^{*}}\left(\Omega'\right)_{s}Y_{l}^{m}\left(\Omega\right)=\delta^{2}\left(\Omega-\Omega'\right)$,
$_{s}Y_{l}^{m^{*}}=\left(-1\right)^{s+m}\,\,\,\,{}_{-s}Y_{l}^{m}$,
and satisfy the spin lowering operation $\mathcal{L}_{\textrm{s}}^{\pm}Y_{\pm s}=\pm\sqrt{\left(l+s\right)\left(l-s+1\right)}Y_{\pm\left(s-1\right)}$.

\section{Green's function of the scalar equation}

The Green's function of the scalar equation can be determined by working
out the solution of the Eq. (\ref{eq:Maxwell Eq1}) with point source:

\begin{equation}
\sin^{2}r\left[\frac{\partial^{2}}{\partial r^{2}}-\frac{\partial^{2}}{\partial\eta^{2}}+\frac{1}{\sin^{2}r}\mathcal{L}_{\textrm{1}}^{+}\mathcal{L}_{\textrm{0}}^{-}\right]G_{0}\left(\eta,\,\, r,\,\,\theta,\,\,\phi\right)=\delta\left(r-r'\right)\delta\left(\eta-\eta'\right)\delta\left(\theta-\theta'\right)\delta\left(\phi-\phi'\right)
\end{equation}

By standard technique, we can make an eigenfunction expansion of the
Green's function\cite{mathew} as 

\begin{equation}
G\left(\eta,\,\, r,\,\,\theta,\,\,\phi:\eta',\,\, r',\,\,\theta',\,\,\phi'\right)=\frac{1}{2\pi}\intop_{-\infty}^{\,\infty}d\omega\,\,\underset{klm}{\sum}\frac{e^{-i\omega\left(\eta-\eta'\right)}}{\omega^{2}-k^{2}}\frac{R_{k}\left(r'\right)R_{k}\left(r\right)}{\sin r\,\,\sin r'}\,\, Y_{l}^{m^{*}}\left(\Omega'\right)\,\,\, Y_{l}^{m}\left(\Omega\right)
\end{equation}

where the radial eighenfunctions are the normalized, appropriately
weighted Gegenbauer polynomials\cite{abram} $R_{k}=N\left(\sin r\right)^{l+1}C_{k-l-1}^{l+1}\left(\cos r\right).$
The addition theorem of spherical harmonics gives

\begin{equation}
\underset{m}{\sum}\,\,\,_{l}Y_{m}^{*}\left(\Omega'\right)\,\,\,_{l}Y_{m}\left(\Omega\right)=\frac{zl+1}{4\pi}C_{l}^{1/2}\left(\cos\beta\right)
\end{equation}
with 
\begin{equation}
\cos\beta=\cos\theta\cos\theta'+\sin\theta\sin\theta'\cos\left(\phi-\phi'\right)
\end{equation}

The $\omega$ integral can be done by the method of residues to give
$\frac{1}{2\pi}\intop_{-\infty}^{\,\infty}dw\frac{e^{-iw\left(\eta-\eta'\right)}}{w^{2}-k^{2}}=-\frac{\sin k\left(\eta-\eta'\right)}{k}$,
with these results we can rewrite the summation as

\begin{eqnarray}
G_{0} & = & -\frac{1}{4\pi}\underset{k=0}{\sum^{\infty}}\underset{l=0}{\sum^{k}}\frac{\sin\left[\left(k+1\right)\left(\eta-\eta'\right)\right]}{\left(k+1\right)}\left(\sin r\sin r'\right)^{l}\nonumber \\
 &  & C_{k-l}^{l+1}\left(\cos r'\right)C_{k-l}^{l+1}\left(\cos r\right)C_{l}^{\frac{1}{2}}\left(\cos\rho\right)\\
 & = & -\frac{1}{4\pi}\underset{k=0}{\sum^{\infty}}\sin\left[\left(k+1\right)\left(\eta-\eta'\right)\right]C_{k}^{(1)}\left(\cos\rho\right)
\end{eqnarray}

where we have used the addition theorem of Gegenbauer polynomial from
reference \cite{grads} and $\cos\rho=\cos r\cos r'+\sin r\sin r'\cos\beta$.

Now 
\begin{eqnarray}
C_{k}^{1}\left(\cos\rho\right) & = & \frac{\sin\left[\left(k+1\right)\rho\right]}{\sin\rho},\,\,\, so\,\, we\,\, find\nonumber \\
G_{0} & =-\frac{1}{4\pi\sin\rho} & \left[\delta\left(\eta-\eta'-\rho\right)-\delta\left(\eta-\eta'+\rho\right)\right]\label{eq:G0}
\end{eqnarray}

representing the retarded and advanced Green's function. Thus, the
solution for the retarded scalar field is 

\begin{equation}
\Phi_{0}\left(\eta,\,\, r,\,\,\theta,\,\,\phi\right)=-\frac{1}{4\pi}\intop_{0}^{\pi}\frac{dr'}{\sin\rho}\int d\Omega'S_{0}\left(\eta-\rho,r',\Omega'\right)
\end{equation}

We can even solve for Lienard-Wiechert like field in closed FRW space
time for a point source moving along the trajectory $\overrightarrow{\xi}\left(\eta'\right)$given
by $S_{0}=\delta\left[\overrightarrow{r'}-\overrightarrow{\xi}\left(\eta'\right)\right]$to
find 
\begin{equation}
\Phi_{0}\left(\eta,\,\, r,\,\,\theta,\,\,\phi\right)=\left.-\frac{1}{4\pi}\frac{1}{\sin\left[\rho\left(\eta'\right)\right]}\frac{1}{\left[1+\frac{d\rho\left(\eta'\right)}{d\eta}\right]}\right|_{\eta'=\eta-\rho\left(\eta'\right)}
\end{equation}

In the limit to flat FRW space time given by $r<<1,\,\,\,\xi<<1,\,\,\sin\rho\sim\rho,$
and $\cos\rho\rightarrow1-\frac{\rho^{2}}{2}\approx\left(1-\frac{r^{2}}{2}\right)\left(1-\frac{\xi^{2}}{2}\right)+r\xi\cos\beta=1-\frac{1}{2}\left|\overrightarrow{r}-\overrightarrow{\xi}\right|^{2}$gives

\begin{equation}
\Phi_{0}\rightarrow\left.-\frac{1}{4\pi}\frac{1}{\left|\overrightarrow{r}-\overrightarrow{\xi}\left(\eta'\right)\right|-\frac{d\overrightarrow{\xi}}{d\eta}.\left(\overrightarrow{r}-\overrightarrow{\xi}\left(\eta'\right)\right)}\right|_{\eta'=\eta-\left|\overrightarrow{r}-\overrightarrow{\xi}\left(\eta'\right)\right|}
\end{equation}

which is the familiar Lienard-Wiechert solution.

\section{Radial Maxwell fields due to Dirac source\label{sec:Radial-Maxwell-fields}}

In this regard, let us solve for the field in the very early universe
when the Dirac particles are copiously produced so that all the available
states are fully occupied. When we sum the source on the left of Eq.
(\ref{eq:Maxwell Eq1}), the second term containing $\mathcal{L}_{\textrm{0}}^{\pm}Y_{\pm\frac{1}{2}}Y_{\mp\frac{1}{2}}^{*}=-\left(l+\frac{1}{2}\right)\left(\left|Y_{\textrm{1/2}}\right|^{2}-\left|Y_{-\textrm{1/2}}\right|^{2}\right)$
will give zero. The consequence is that there is no source for the
radial magnetic field. This is general in that whenever the $\pm m$
states are paired, their contributions cancel out. In this case, there
is an isotropic source of electric field only as $\underset{m}{\sum}\left(\left|Y_{\textrm{1/2}}\right|^{2}+\left|Y_{-\textrm{1/2}}\right|^{2}\right)=\frac{2\left(2l+1\right)}{4\pi}$
. Thus the Maxwell Eq. (\ref{eq:Maxwell Eq2}) reduces to 

\begin{eqnarray}
\mathsf{\mathsf{\mathcal{D}}}_{0}^{\mp}\left[\sin^{2}r\, F_{r\eta}\right] & = & \underset{kl}{\sum}\left|\phi_{\pm\frac{1}{2}}\right|^{2}\frac{2\left(2l+1\right)}{4\pi}\\
or,\,\,\,\,\,\,\,\,\,\,\,\,\,\,\,\,\,\,\frac{\partial}{\partial r}\sin^{2}r\,\, F_{r\eta} & = & \underset{kl}{\sum}\frac{\left(2l+1\right)}{4\pi}\left(\left|\phi_{\frac{1}{2}}\right|^{2}+\left|\phi_{-\frac{1}{2}}\right|^{2}\right)\\
and\,\,\,\,\,\,\,\,\,\,\,\,\,\,\,\,\,\,\frac{\partial}{\partial r}\sin^{2}r\,\, F_{r\eta} & = & \underset{kl}{\sum}\frac{\left(2l+1\right)}{4\pi}\left(\left|\phi_{\frac{1}{2}}\right|^{2}-\left|\phi_{-\frac{1}{2}}\right|^{2}\right)
\end{eqnarray}

The solution is 
\begin{eqnarray}
4\pi\sin^{2}r\, F_{r\eta} & = & \frac{1}{2}\underset{kl}{\sum}\left(2l+1\right)\intop_{0}^{\,\,\, r}dr'\left(\left|Z_{+}\right|^{2}+\left|Z_{-}\right|^{2}\right)\\
 & = & \underset{k=0}{\sum^{\infty}}\underset{l=0}{\sum^{k}}\intop_{0}^{\,\,\, r}dr\sin r\,\,(1-\cos r)^{l+\frac{1}{2}}(1+\cos r)^{l-\frac{1}{2}}\nonumber \\
 &  & \left[\left\{ P_{k-l}^{\left(l+\frac{1}{2},l+\frac{1}{2}\right)}\left(\cos r\right)\right\} ^{2}\left|T_{+}\right|^{2}+\left\{ P_{k-l}^{\left(l+\frac{1}{2},l+\frac{1}{2}\right)}\left(\cos r\right)\right\} ^{2}\left|T_{-}\right|^{2}\right],\nonumber \\
\end{eqnarray}

where $P_{n}^{\left(\alpha,\beta\right)}$are appropriately weighted
and normalized Jacobi polynomials, is just a statement of Gauss law,
as the right handside is the co-moving number of enclosed particles.
Here to first order WKB, $\left|T_{\pm}\right|^{2}=1\pm\frac{aM}{\sqrt{\left(k+\frac{3}{2}\right)^{2}+a^{2}M^{2}}}$.
The field intensity $F_{r\eta}$are shown in Fig (\ref{fig: F vs r for different masses})
for differnt values of M and in Fig. (\ref{fig: F vs r for different k})
for some values of the comoving momentum $k$ of the Dirac particles.We
take those values to be representative of the proton and electron.
At any finite time after the big bang, we see that the protons produce
a stronger field than the electron. The two fields do not have the
same magnitude even for the lowest state $k=0$, and the heavier mass
consistently produces stronger field of higher momentum also. These
are exact solutions of the field equations, and we feel that this
effect will have strong consequence on the formation of hydrogen and
other atoms. 

Next, we discuss the solutions in the case after particle-antiparticle
anhilation when the electron number becomes drastically reduced. Then
there will not always be pairing of the $\pm m$ states. For this,
we can use the well known generalized Clebsch Gordon expansion

\begin{eqnarray}
_{s_{1}}Y_{l_{1}\,\,\,\, s_{2}}^{m_{1}}Y_{l_{2}}^{m_{2}} & = & \underset{l=\left|l_{1}-l_{2}\right|}{\sum^{l_{1}+l_{2}}}\sqrt{\frac{\left(2l_{1}+1\right)\left(2l_{2}+1\right)\left(2l+1\right)}{4\pi}}\left(\begin{array}{ccc}
l_{1} & l_{2} & l\\
m_{1} & m_{2} & -\left(m_{1}+m_{2}\right)
\end{array}\right)\nonumber \\
 &  & \left(\begin{array}{ccc}
l_{1} & l_{2} & l\\
-s_{1\,\,} & -s_{2} & \,\,\,\,\, s_{1}+s_{2}
\end{array}\right)\,\,\,\,\,_{s_{1}+s_{2}}Y_{l}^{m_{1}+m_{2}}
\end{eqnarray}

where $\left(\begin{array}{ccc}
j{}_{1} & j_{2} & j\\
m_{1} & m_{2} & m
\end{array}\right)$ are the wigner 3-j symbol. Hence, we find

\begin{eqnarray}
\left|Y_{+}\right|^{2}+\left|Y_{-}\right|^{2} & = & 2\left(-1\right)^{m-\frac{1}{2}}\underset{l=0}{\sum^{l-\frac{1}{2}}}\sqrt{\frac{\left(2l+1\right)^{2}\left(4l+1\right)}{4\pi}}\nonumber \\
 &  & \left(\begin{array}{ccc}
l & l & 2L\\
m & -m & 0
\end{array}\right)\left(\begin{array}{ccc}
l & l & 2L\\
\frac{1}{2} & -\frac{1}{2} & 0
\end{array}\right)\,\,\,_{0}Y_{2L}^{0}
\end{eqnarray}

contributes only the even angular momentum states. Similarly,

\begin{eqnarray}
\mathcal{L}_{\textrm{1}}^{\pm}Y_{\pm\frac{1}{2}}Y_{\mp\frac{1}{2}}^{*} & = & 2\left(l+\frac{1}{2}\right)\left(-1\right)^{m-\frac{1}{2}}\underset{l=0}{\sum^{l-\frac{1}{2}}}\sqrt{\frac{\left(2l+1\right)^{2}\left(4l+3\right)}{4\pi}}\nonumber \\
 &  & \left(\begin{array}{ccc}
l & l & 2L+1\\
m & -m & 0
\end{array}\right)\left(\begin{array}{ccc}
l & l & 2L+1\\
\frac{1}{2} & -\frac{1}{2} & 0
\end{array}\right)\,\,\,_{0}Y_{2L}^{0}
\end{eqnarray}

contributes the odd states. So, we can conclude that the radial electric
field is composed of even angular momentum states and the odd ones
compose radial magnetic field. These solutions and the solutions for
$\phi_{\pm1}$ will be considered in future work.

\section{Conclusion}

We have solved some of the Maxwell equations in FRW spacetime with
the source given by Dirac field. The variables are all separable.
For the scalar field representing the components $F_{r\eta}$ and
$F_{\theta\phi}$, we are able to solve the Green's function which
reduces to the familiar ones of electrodynamics in the appropriate
limit to flat case. The retarded part is used to derive the Lienard
Wiechert type solution for moving point source.

In section \ref{sec:Radial-Maxwell-fields}, we consider the source
given by Dirac potentials. In the case that all the available states
of the Dirac particle are filled, the magnetic field $F_{\theta\phi}$
is found to vanish. When we compare the electric filed strength $F_{r\eta}$
due to Dirac particles of mass ratio 10000, they are somewhat different
with the lighter one contributing more. At this lowest level the fields
due to proton and electron are not exactly opposite of each other.
This should have observable effects on the formation of atoms and
particularly on the time of decoupling. When we consider individual
azimuthal states of Dirac particles, the source term vanishes. When
we consider Dirac particles after the particle-antiparticle annihilation,
the number density is greatly decreased so that all the available
states cannot be occupied. In particular, if all the $\pm m$ azimuthal
states are not paired, then the source for $F_{r\eta}$ and $F_{\theta\phi}$
do not vanish. In that case $F_{r\eta}$ is composed of even angular
momentum ($l$) states and $F_{\theta\phi}$ is composed of odd states.
So, there is the possibility of the existence of a primordial magnetic
field generated just after $e^{+}-e^{-}$ annihilation.

\end{document}